\documentclass[aps,showkeys,showpacs,prd,superscriptaddress,twocolumn]{revtex4}
\usepackage[dvips]{graphicx}
\usepackage{graphicx}
\usepackage{dcolumn}
\usepackage{bm}
\usepackage{epsfig}
\usepackage{latexsym,bm}

\begin{document}

\title{Yangian description for decays and possible explanation of $X$ in the decay $K^0_L\rightarrow \pi^0 \pi^0 X$}
\author{Li-Jun Tian}%
\email{tianlijun@shu.edu.cn}%
\affiliation{Department of Physics,%
Shanghai University, Shanghai, 200444, China}%
\affiliation{Shanghai Key Lab for Astrophysics,
Shanghai, 200234, China}%
\author{Yan-Ling Jin}%
\email{jinyanling@shu.edu.cn}%
\affiliation{Department of Physics,%
Shanghai University, Shanghai, 200444, China}%
\affiliation{Shanghai Key Lab for Astrophysics,
Shanghai, 200234, China}%
\author{Ying Jiang}
\email{yjiang@shu.edu.cn}
\affiliation{Department of Physics,%
Shanghai University, Shanghai, 200444, China}%
\affiliation{Shanghai Key Lab for Astrophysics,
Shanghai, 200234, China}%

\begin{abstract}

In this letter, hadronic decay channels of light pseudoscalar
mesons are realized in Yangian algebra. In the framework of
Yangian, we find that these decay channels can be formulated by
acting transition operators, composed of the generators of
Yangian, on the corresponding pseudoscalar mesons. This new
description of decays allows us to present a possible
interpretation of the new unknown particle $X$ in the decay
$K^0_L\rightarrow \pi^0 \pi^0 X$: it is an entangled state of
$\pi^0$ and $\eta$.
\end{abstract}


\maketitle \baselineskip=12pt

Particle decay is one of the most fascinating phenomena in the
physics world, it also plays a significant role in searching for
new particles, which is always one of the main tasks of
fundamental research.

Recently, HyperCP collaboration has reported \cite{park-1} the
first evidence for the decay $\Sigma^+ \rightarrow p\mu^+\mu^-$
with branching ratio at the level of $10^{-8}$ from data taken by
the HyperCP (E871) experiment at Fermilab. Narrow range of dimuon
masses in three observed events in that experiment suggested a new
particle $X$ with mass of 214.3 MeV. Stimulated by this exciting
new discovery, some important literatures on interpreting and
finding this new particle came out: Gorbunov and Rubakov suggested
\cite{gorbunov-1} that Sgoldino - superpartner of goldino - may
take the responsibility for the anomalous events mentioned above;
X.-G. He and his colleague \cite{he-1} argued that this new
"HyperCP particle" can be identified with the light pseudoscalar
Higgs boson in the next-to-minimal supersymmetric standard model.
The most recent effort in finding this novel particle $X$ was done
by Tung {\it et al.} \cite{Tung}, they searched for the decay of
$K_L^0 \rightarrow \pi^0 \pi^0 X$, $X\rightarrow \gamma \gamma$
with the E391a detector at KEK, and reported an upper limit of the
branching ratio at the level of $10^{-7}$, and the possible mass
region of this new pseudoscalar particle is 194.3-219.3 MeV.

In this letter, we will probe the issue of meson decays via
Yangian algebra $Y(su(3))$. In fact, investigating the hadronic
decay channels of light pseudoscalar mesons via Yangian is an
important and powerful approach. In the framework of Yangian, we
find that these decay channels can be formulated by acting
transition operators, composed of the generators of Yangian, on
the corresponding pseudoscalar mesons. This new description of
decays allows us to present a possible interpretation of the new
unknown particle $X$ in the decay $K^0_L\rightarrow \pi^0 \pi^0
X$, and we show that $X$ might be an entangled state of $\pi^0$
and $\eta$.

As is known, mesons can be looked upon as entanglement states of
quarks, and the high energy quantum teleportation related to kaons
has already been under investigation \cite{Shi1}. As a co-product,
the entanglement degrees of initial and final states of the
transition processes are also presented in this letter.

{\it Yangian $Y(su(3))$} -- Yangian algebras, related to a
rational solution of a classical Yang-Baxter equation, were
presented by Drinfeld\cite{drinfeld} in 1985. Many models have
been demonstrated to possess Yangian symmetry, such as
Calogero-Sutherland model\cite{Hikami}, the one-dimensional
Hubbard model\cite{Inozemtsev}, long-range $gl(N)$ integrable spin
chains\cite{Niklas} and so on. Not only can Yangian algebra
describe the symmetry of quantum integrable models, but also can
present the transitions of the states\cite{Ge1} between different
weights beyond the Lie algebra. Our main result in this letter are
exactly based on this important and particular transition effect.

The Yangian related to the Lie algebra $su(3)$ is denoted as
$Y(su(3))$\cite{Ge2}. It is generated by the generators
$\{{I}_{\alpha},{J}_{\alpha}\}$ which are usually defined as
follows,
\begin{eqnarray}
&&I^a=\sum_iF_i^a,\nonumber\\
&&J^a=\mu{I_1^a}+\nu{I_2^a}+\frac{i}{2}\lambda{f_{abc}\sum_{i{\neq}j}{\omega}_{ij}I_i^bI_j^c}
\end{eqnarray}
with $i,j=1,2$. Here
\begin{eqnarray}
\label{omega}{\omega}_{ij}=\left\{
\begin{array}{l}
1\;\;\;\;\;\;\;\;i{>}j\\
-1\;\;\;\;\;i{<}j\\
0\;\;\;\;\;\;\;\;i{=}j
\end{array} \right.,
\end{eqnarray}
$\mu$, $\nu$, $\lambda$ are parameters or Casimir operators, and
$f_{abc}$ $(a,b,c=1,2,3)$ are the structure constants of $su(3)$
algebra. $\{F_{i}^{a}, a=1,2,\cdots,8\}$ form a local $su(3)$ on the
$i$ site, and are equal to half of the corresponding Gell-Mann
matrices.

For brevity, transformations are introduced as follows:
$\bar{I}^{\pm}=J^{1}{\pm}iJ^{2}, \bar{U}^{\pm}=J^{6}{\pm}iJ^{7},
\bar{V}^{\pm}=J^{4}{\pm}iJ^{5}, \bar{I}^{3}=J^{3},
\bar{I}^{8}=\frac{2}{\sqrt{3}}J^{8}$. Different combinations of
these operators will provide us with different transition
operators. We will show that hadronic decay channels of
pseudoscalar mesons can be formulated by acting properly chosen
transition operators on the corresponding initial states of
mesons. In order to illustrate this clearly, in the following, we
will discuss different cases one by one.

{\it Yangian $Y(su(3))$ in $\eta$ decay channels} -- The $\eta$
and $\eta^{'}$ mesons play a special role in understanding low
energy QCD. They are isoscalar members of the nonet of the
lightest pseudoscalar mesons. and $\eta$-$\eta'$ mixing system is
one of the most attractive problems all
along\cite{Fritzsch,Isgur,Fazio,Escribano,Oset,Borasoy}. Moreover,
the decays of $\eta$ provide information about the pseudoscalar
form factor\cite{Kupsc}.

We choose $\eta$, which is superposition of singlet and octet of
$su(3)$, as the initial state
\begin{eqnarray}
\label{1}
|\eta\rangle_{ini}=\alpha_1|\eta^{0'}\rangle+\alpha_2|\eta^0\rangle,
\end{eqnarray}
where $\alpha_1$ and $\alpha_2$ are the normalized real amplitudes
and they satisfy $\alpha_1^2+\alpha_2^2=1$.
$|\eta^{0'}\rangle=\frac{\sqrt3}{3}(|u\bar{u}\rangle+|d\bar{d}\rangle+|s\bar{s}\rangle)$,
$|\eta^0\rangle=\frac{\sqrt6}{6}(-|u\bar{u}\rangle-|d\bar{d}\rangle+2|s\bar{s}\rangle)$.

\begin{figure}[h]
\includegraphics[angle=0,width=8cm]{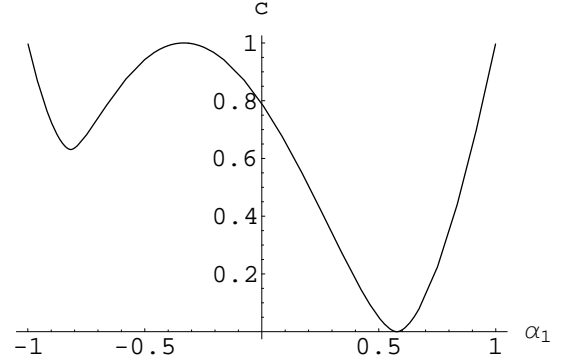}
 \caption{The entanglement degree of the initial state $|\eta\rangle_{ini}$ varies with the amplitude of the singlet state.}
\label{fig1}
\end{figure}

As is known, the entanglement degree of the genuine N-particle
qutrit pure state\cite{Pan} is measured by the mean entropy as
follows \cite{wootters}
\begin{eqnarray}
\label{2} C^{(N)}_\Phi=\left\{
\begin{array}{l}
\frac1N\sum_{i=1}^nS_{(i)}\;\;\;\;\;$if$\;S_i\neq0\;\forall\;i\\
0\;\;\;\;\;\;\;\;\;\;\;\;\;\;\;\;\;\;\;\;\;\;$otherwise$
\end{array} \right.,
\end{eqnarray}
where $S_i=-Tr((\rho_\Phi)_iLog_3(\rho_\Phi)_i)$ is the reduced
partial Von Neumann entropy for the $i$th particle only, with the
other N-1 particles traced out, and $(\rho_\Phi)_i$ is the
corresponding reduced density matrix. The system we discuss here
is bipartite qutrit, N=2, thus the entanglement degree of the
initial state can be gotten by applying the Eq.~(\ref{2}),
\begin{eqnarray}
C_{ini}=-2(\frac{\sqrt3}{3}\alpha_1-\frac{\sqrt6}{6}\alpha_2)^2Log_3(\frac{\sqrt3}{3}\alpha_1
-\frac{\sqrt6}{6}\alpha_2)^2\nonumber\\
-(\frac{\sqrt3}{3}\alpha_1+\frac{\sqrt6}{3}\alpha_2)^2Log_3(\frac{\sqrt3}{3}\alpha_1+\frac{\sqrt6}{3}\alpha_2)^2.
\end{eqnarray}
 Fig.~\ref{fig1} shows the variation of entanglement degree of initial state depending on the
 amplitude of the singlet state.

Now, let us take the transition operator $P=3{\bar{I}^{3}}$ and
act it on the initial state $|\eta\rangle_{ini}$, we then get the
final state
\begin{eqnarray}
|\eta\rangle_{fin}=P|\eta\rangle_{ini}=|\pi^0\rangle+|\pi^0\rangle+|\pi^0\rangle,
\end{eqnarray}
with the normalization condition
$(\mu+\nu)(\alpha_1-\frac{\sqrt2}{2}\alpha_2)=\pm\sqrt6$.
Calculation shows that the entanglement degree of the final state
$C_{fin}$ is a constant $C_{fin}=0.631$, independent with the
amplitude $\alpha_i$ of the initial state.

From the above formula, to our surprise, we find that this
transition is exactly corresponding to the decay channel of
$\eta\rightarrow\pi^0\pi^0\pi^0$!

Other hadronic decay channels of $\eta$ can be also obtained with
the same method in the framework of Yangian. For instance, after
tedious calculation, we find that, by taking
$P=\sqrt6\bar{I}^3-\sqrt3(\bar{I}^{+}-\bar{I}^{-})$ with
$\nu=-\frac{\lambda}{2}$ and
$(\mu-\frac{\lambda}{2})(\alpha_1-\frac{\sqrt2}{2}\alpha_2)=\pm1$
or $P=\sqrt6\bar{I}^3+\sqrt3(\bar{I}^{+}-\bar{I}^{-})$ with
$\mu=\frac{\lambda}{2}$ and
$(\nu+\frac{\lambda}{2})(\alpha_1-\frac{\sqrt2}{2}\alpha_2)=\pm1$
, the initial state $|\eta\rangle_{ini}$  transits to
\begin{eqnarray}
|\eta\rangle_{fin}=|\pi^0\rangle+|\pi^{+}\rangle+|\pi^{-}\rangle
\end{eqnarray}
with entanglement degree of $C_{fin}=0.118$. This corresponds to
the decay channel of $\eta\rightarrow\pi^0\pi^{+}\pi^{-}$.

Moreover, the final state becomes
\begin{eqnarray}
|\eta\rangle_{fin}=|\pi^{+}\rangle+|\pi^{-}\rangle,
\end{eqnarray}
when the transition operator takes the form of
$P=\sqrt2(\bar{I}^{+}-\bar{I}^{-})$ with normalizing condition
$(\mu+\nu)(\alpha_1-\frac{\sqrt2}{2}\alpha_2)=\pm\frac{\sqrt6}{2}$.
Apparently, this is just the decay channel of
$\eta\rightarrow\pi^{+}\pi^{-}$.

The results mentioned above are summarized in table~\ref{table1}.

\begin{table}
\begin{center}
\caption{Yangian $Y(su(3))$ in $\eta$ decay channels}
\begin{tabular}{|c|c|c|c|c|c|} \hline
&&$normal-$&&&\\
$|\eta\rangle_{ini}$&$P$&izing&$|\eta\rangle_{fin}$&$C_{fin}$&$decay$\\
&&condition&&&\\ \hline\hline
&           &$(\mu+\nu)(\alpha_1$        &$|\pi^0\rangle$                 &&$\eta$\\
&$3\bar{I}^3$&$-\frac{\sqrt2}{2}\alpha_2)$&$+|\pi^0\rangle$           &0.631&$\rightarrow\pi^0$\\
&           &$=\pm\sqrt6$                &$+|\pi^0\rangle$                &&$\pi^0\pi^0$\\
\cline{2-2}\cline{3-3}\cline{4-4}\cline{5-5}\cline{6-6}

               &                             &$(\mu-\frac\lambda2)(\alpha_1$&                                &        &\\
               &$\sqrt6\bar{I}^3-$           &$-\frac{\sqrt2}{2}\alpha_2)$  &                                &        &\\
               &$\sqrt3(\bar{I}^+-\bar{I}^-)$&$=\pm1\;\;and$                &                                &        &\\
$|\eta\rangle_{ini}=$&                             &$\nu=-\frac\lambda2$          &$|\pi^0\rangle$                 &0.118   &$\eta$\\
\cline{2-2}\cline{3-3}

$\alpha_1|\eta^{0'}\rangle$ &                             &$(\nu+\frac\lambda2)(\alpha_1$&$+|\pi^+\rangle$ &&$\rightarrow\pi^0$\\
$+\alpha_2|\eta^0\rangle$   &$\sqrt6\bar{I}^3+$           &$-\frac{\sqrt2}{2}\alpha_2)$  &$+|\pi^-\rangle$ &&$\pi^+\pi^-$\\
                            &$\sqrt3(\bar{I}^+-\bar{I}^-)$&$=\pm1\;\;and$                &                 &&\\
                            &                             &$\mu=\frac\lambda2$           &                 &&\\
\cline{2-2}\cline{3-3}\cline{4-4}\cline{5-5}\cline{6-6}

&                      &$(\mu+\nu)(\alpha_1$        &                                &&$\eta$\\
&$\sqrt2(\bar{I}^+-\bar{I}^-)$ &$-\frac{\sqrt2}{2}\alpha_2)$&$|\pi^+\rangle$   &0.631&$\rightarrow\pi^+$\\
&                      &$=\pm\frac{\sqrt6}{2}$       &$+|\pi^-\rangle$                &&$\pi^-$\\
\hline
\end{tabular}
\label{table1}
\end{center}
\end{table}

{\it Yangian $Y(su(3))$ in $\eta^{'}$ decay channels} --
  $\eta^{'}$ is the most
esoteric meson of the pseudoscalar nonet, closely related to the
axial U(1) anomaly\cite{Kupsc}. We take the initial state
$|\eta^{'}\rangle_{ini}$ which is also the superposition of $su(3)$
singlet and octet
\begin{eqnarray}
|\eta^{'}\rangle_{ini}=\frac{1}{3}|\eta^{0'}\rangle-\frac{2\sqrt2}{3}|\eta^0\rangle
\end{eqnarray}
with entanglement degree $C_{ini}=0.790$.

Here we choose the transition operators as
$P_\pm=\sqrt3\bar{I}^8\pm\frac{3}{\mu-\nu-\lambda}(\bar{I}^+-\bar{I}^-)$
and act them on the initial state $|\eta^{'}\rangle_{ini}$,
respectively, we get the same final state
\begin{eqnarray}
|\eta^{'}\rangle_{fin}=|\eta\rangle+|\pi^+\rangle+|\pi^-\rangle.
\end{eqnarray}
Its entanglement degree is $C_{fin}=0.973$. Here the normalization
condition reads $\mu+\nu=\pm\frac{3\sqrt3}{2}$. Clearly, we find
that this transition corresponds to the decay channel of
$\eta^{'}\rightarrow\pi^+\pi^-\eta$.

  Acting the operators
$P_\pm=\sqrt5\bar{I}^3\pm\frac{\sqrt15}{4}\bar{I}^8$ on
$|\eta^{'}\rangle_{ini}$, we get correspondingly the final state
as
\begin{eqnarray}
|\eta^{'}\rangle_{fin}=|\eta\rangle+|\pi^0\rangle+|\pi^0\rangle
\end{eqnarray}
with the normalization condition $\mu+\nu=\pm\frac{2\sqrt30}{5}$.
And the entanglement degree of the final state is $C_{fin}=0.786$.
This provides us with another decay channel
$\eta^{'}\rightarrow\pi^0\pi^0\eta$.

These two cases of $\eta^{'}$ decay channels are listed in detail
in table~\ref{table2}.

\begin{table}
\begin{center}
\caption{Yangian $Y(su(3))$ in $\eta^{'}$ decay channels}
\begin{tabular}{|c|c|c|c|c|c|} \hline
&&$normal-$&&&\\
$|\eta^{'}\rangle_{ini}$&$P_\pm$&izing&$|\eta^{'}\rangle_{fin}$&$C_{fin}$&$decay$\\
&&condition&&&\\ \hline\hline

                     &$\sqrt3\bar{I}^8\pm$            &                         &$|\eta\rangle$                  &                     &$\eta^{'}$\\
                     &$\frac{3}{\mu-\nu-\lambda}$     &  $\mu+\nu$              &$+|\pi^+\rangle$                &0.973               &$\rightarrow\eta$\\
                     &$(\bar{I}^+-$                   &   $=\frac{3\sqrt3}{2}$  &$+|\pi^-\rangle $               &                    &$\pi^+\pi^-$\\
$|\eta^{'}\rangle_{ini}=$  &$\bar{I}^-)$                    &                         &                                &                    &\\
\cline{2-2}\cline{3-3}\cline{4-4}\cline{5-5}\cline{6-6}

$\frac13|\eta^{0'}\rangle-$      &                                &                                          & $|\eta\rangle$                   &     &$\eta^{'}$\\
$\frac{2\sqrt2}{3}|\eta^0\rangle$& $\sqrt5\bar{I}^3\pm$           & $\mu+\nu$                                &  $+|\pi^0\rangle$                &0.786&$\rightarrow\eta$\\
                                 & $\frac{\sqrt15}{4}\bar{I}^8$   & $=\pm\frac{2\sqrt30}{5}$                 &    $+|\pi^0\rangle$              &     &
                                 $\pi^0\pi^0$\\\hline

\end{tabular}
\label{table2}
\end{center}
\end{table}

{\it Yangian $Y(su(3))$ in $\pi^\pm$ and $\kappa_{L(S)}^0$ decay
channels} -- By making use of the same approach, the hadronic
decay channels of $\pi^\pm$ and $\kappa_{L(S)}^0$ can also be
realized in Yangian algebra based on its transition effect. The
results are summarized in table~\ref{table3}.

\begin{table}
\begin{center}
\caption{Yangian $Y(su(3))$ in $\pi^\pm$ and $\kappa_{L(S)}^0$ decay
channels}
\begin{tabular}{|c|c|c|c|c|c|} \hline
&&$normal-$&&&\\
$|\varphi\rangle_{ini}$&$P$&izing&$|\varphi^{'}\rangle_{fin}$&$C_{fin}$&$decay$\\
&&condition&&&\\ \hline\hline

                  &           &                             &$|\pi^+\rangle$                 &                              &$\kappa^+$\\
$|\kappa^+\rangle$&           &                             &$+|\pi^+\rangle$                 &                              &$\rightarrow\pi^+$\\
                  &$\sqrt5\bar{U}^+$&$\lambda=\pm\frac{2}{\sqrt5}$&$+|\pi^-\rangle$                &0.455                        &$\pi^+\pi^-$\\
\cline{1-1}\cline{4-4}\cline{6-6}
                  &           &                             &$|\pi^-\rangle$                 &                              &$\kappa^-$\\
$|\kappa^-\rangle$&           &                             &$+|\pi^-\rangle$                 &                              &$\rightarrow\pi^-$\\
                  &           &                             &$+|\pi^+\rangle$                &                              &$\pi^-\pi^+$\\ \hline

$|\kappa_L^0\rangle$             &                                                               &                                             &                                &        &$\kappa_L^0$\\
$(|\kappa_S^0\rangle)=$          &$\sqrt2\bar{V}^+$                                              &$\lambda=\pm\frac{\sqrt2}{\alpha_1+\alpha_2}$&$|\pi^+\rangle$                 &0.631&$(\kappa_S^0)$\\
$\alpha_1|\kappa^0\rangle$       &                                                               &                                             &$+|\pi^-\rangle$                &        &$\rightarrow$\\
$+\alpha_2|\bar{\kappa}^0\rangle$&                                                               &                                             &                                &        &$\pi^+\pi^-$\\
\hline
\end{tabular}
\label{table3}
\end{center}
\end{table}

From the above, we see that hadronic decay channels of light
pseudoscalar mesons can be realized in Yangian algebra. In the
framework of Yangian, we find that these decay channels can be
formulated by acting transition operators - composed of the
generators of Yangian - on the corresponding pseudoscalar mesons.
This new description of decays allows us to present a possible
interpretation of the new unknown particle $X$ in the decay
$K^0_L\rightarrow \pi^0 \pi^0 X$, as will be discussed below.

{\it $X$ in $\kappa_L^0$ decay channel} -- Several months ago, Y.
C. Tung et al. \cite{Tung} searched for a new light pseudo-scalar
particle $X$ - first evidence of its existence was reported by the
HyperCP collaboration in 2005 \cite{park-1} - in a decay channel
of $\kappa_L^0$
\begin{eqnarray}
\kappa_L^0\rightarrow\pi^0\pi^0X \label{eq12}
\end{eqnarray}
with the E391a detector at KEK, and reported the branching ratio
for this channel being about $10^{-7}$ at the 90\% confidence
level. It is reported \cite{park-1, Tung} that the mass of this
unknown pseudo-scalar particle is about $214.3 MeV$, and $X$
immediately decays to two photons.

With the help of the method mentioned above, we would like to
provide a possible interpretation of this unknown particle $X$.

We consider the initial state of
\begin{eqnarray}
|\kappa_L^0\rangle_{ini}=\alpha_1|\kappa^0\rangle+\alpha_2|\bar{\kappa}^0\rangle.
\end{eqnarray}
The transition operator which are going to be acted on this
initial state takes the form of
$P=\bar{U}^++(0.131c-0.500)\bar{U}^-$ with $c$ a tunable
parameter. Associated with the normalization conditions
$|8.027-0.707c|^2+|c|^2=1$, $\mu=-\nu$ and
$(\nu-\frac12\lambda)(\alpha_1-\alpha_2)=-4.100$ , we find that a
surprising result pops up
\begin{eqnarray}
|\kappa_L^0\rangle_{fin}=\pi^0+\pi^0+(0.899\pi^0+0.438\eta).
\end{eqnarray}
By comparing this formula with Eq.~(\ref{eq12}), it is obvious
that the unknown particle $X$ could be interpreted as an entangled
state of $\pi^0$ and $\eta$ with a certain proportion. With the
masses of $\pi^0$ and $\eta$ being $M_{\pi^0}\approx135MeV/c^2$
and $M_\eta\approx548MeV/c^2$, we have immediately the mass of the
particle $X$ being
$M_X=(0.899^2*135+0.438^2*548)MeV/c^2\approx214.3MeV/c^2$ which is
in agreement with the experimental observation \cite{park-1,Tung}.
In addition, both $\pi^0$ and $\eta$ can decay to $2\gamma$, this
is again coincident with experiments.

In summary, we find that the hadronic decay channels of
pseudo-scalar mesons can be reformulated under the framework of
Yangian by acting properly chosen transition operators -
consisting of generators of Yangian $Y(su(3))$ - on corresponding
initial states.

Besides conventional theories and methods in investigating the
hadronic decay channels of pseudo-scalar mesons, Yangian shows us
another way to look insight into the phenomena of meson decay, and
provides us with an alternative method to investigate exotic
effects observed in experiments. One of the examples is the
puzzling new particle $X$ in a new decay channel of
$\kappa_L^0\rightarrow \pi^0\pi^0 X$, our result indicates that
this new particle $X$ might be interpreted as an entangled state
of $\pi^0$ and $\eta$.

{\it Acknowledgement} --
  This work is in part supported by the NSF of China under Grant
  No.10775092 and No.10845002, Shanghai Leading Academic Discipline
  Project (Project number S30105) and Shanghai Research Foundation
  No.07d222020. Financial support from the Science and Technology
  Committee of Shanghai Municipality under Grant Nos. 08dj1400202
  and 09PJ1404700 is also acknowledged.


\begin{thebibliography}{99}

\bibitem{park-1}H.K. Park, {\it et al.}, Phys. Rev. Lett. {\bf
94}, 021801 (2005)

\bibitem{gorbunov-1}D.S. Gorbunov, and V.A. Rubakov, Phys. Rev.
D{\bf 73}, 035002 (2006)

\bibitem{he-1}X.G. He, J. Tandean, and G. Valencia, Phys. Rev.
Lett. {\bf 98}, 081802 (2007)

\bibitem{Tung}Y. C. Tung et al., Phys. Rev. Lett.
{\bf 102}, 051802 (2009)

\bibitem{Shi1}Y. Shi, Phys. Lett. B {\bf{641}} 75 (2006).

\bibitem{drinfeld} V. G. Drinfeld, Sov. Math. Dokl. {\bf{32}} 254
(1985).

\bibitem{Hikami}K. Hikami and M. Wadati, J. Phys. Soc. Japan {\bf{62}} 469 (1993).

\bibitem{Inozemtsev}F. G\"{o}mann and V. Inozemtsev, Phys. Lett. A {\bf{214}} 161 (1996).

\bibitem{Niklas} N. Beisert and D. Erkal, J. Stat. Mech P03001 (2008).

\bibitem{Ge1}M. L. Ge, L. C. Kwek, C. H. Oh and K. Xue, Czechoslovak J. Phys.
{\bf{50}} 1229 (2000).

\bibitem{Ge2}M. L. Ge and K. Xue, Yang$-$Baxter Equation (Shanghai Scientific and Technical Publishers, 1999).

\bibitem{Oset}E. Oset, J. R. Pel\'{a}ez and L. Roca, AIP Conf. Proc.
{\bf{717}} 185 (2004).

\bibitem{Borasoy} B. Borasoy and R. Ni{\ss}ler, Eur. Phys. J. A
{\bf{33}} 95 (2007).

\bibitem{Fritzsch} H. Fritzsch and J. D. Jackson, Phys. Lett. B {\bf{66}} 365 (1977).

\bibitem{Isgur}N. Isgur, Phys. Rev. D {\bf{13}} 122 (1976).

\bibitem{Fazio}F. D. Fazio and M. R. Pennington, J. High Energy
Phys. {\bf{07}} 051 (2000).

\bibitem{Escribano}R. Escribano and J. Nadal,  J. High Energy
Phys. {\bf{05}} 006 (2007).

\bibitem{Kupsc}A. Kup$\acute{s}\acute{c}$, AIP Conf. Proc. {\bf{950}} 165 (2007).

\bibitem{Pan}F. Pan, G. Y. Lu and J. P. Draayer, Inter. J. Mod. Phys. {\bf{20}} 1333 (2006).

\bibitem{wootters}S. Hill, W.K. Wootters, Phys. Rev. Lett. {\bf
78}, 5022 (1997); W.K. Wootters, Phys. Rev. Lett. {\bf 80}, 2245
(1998)



\end{thebibliography}
\end{document}